\documentclass[12pt]{article}
\usepackage{ifthen}
\newcommand\haveamsmath{1}
\ifthenelse{\haveamsmath=1}{%
\usepackage{amsmath,bm,ifthen}}
{\usepackage{amsfonts,amssymb}}

\renewcommand{\baselinestretch}{1.4}
\def\be{\begin{equation}}
\def\ee{\end{equation}}
\def\bear{\begin{eqnarray}}
\def\eear{\end{eqnarray}}

\def\bS{\mathbf{S}}

\def\bi{\bibitem}

\newcommand{\x}{{\vec x}}
\renewcommand{\P}{{\vec P}}
\newcommand{\<}{\langle}
\renewcommand{\>}{\rangle}
\newcommand{\ti}{{t_{\mathrm{i}}}}
\newcommand{\tf}{{t_{\mathrm{f}}}}
\renewcommand{\Re}{{\mathrm{Re}\,}}
\renewcommand{\Im}{{\mathrm{Im}\,}}

\usepackage{graphicx}
\def\ukrp{u_{k,R,+}}
\def\ukrm{u_{k,R,-}}
\def\uklp{u_{k,L,+}}
\def\uklm{u_{k,L,-}} 

\begin{document}

\begin{titlepage}

\begin{flushright}
hep-th/0212072\\
NSF-ITP-02-175\\
INT-PUB 02-53
\end{flushright}
\vfil

\begin{center}
{\huge  
Schwinger-Keldysh Propagators from}\\
\vspace{3mm}
{\huge AdS/CFT Correspondence}\\
\end{center}

\vfil
\begin{center}
{\large C. P. Herzog$^*$ and D. T. Son$^\dagger$}\\
\vspace{5mm}
$*$ Kavli Institute for Theoretical Physics,\\
University of California, Santa Barbara, CA  93106, USA\\
{\tt herzog@kitp.ucsb.edu}\\
\vspace{3mm}
$\dagger$ Institute for Nuclear Theory, \\
University of Washington, Seattle, WA  98195, USA \\
{\tt son@phys.washington.edu}\\

\vspace{3mm}
\end{center}

\vfil

\begin{center}

{\large Abstract}
\end{center}

\noindent
We demonstrate how to compute real-time Green's functions for a 
class of finite temperature field theories from their AdS
gravity duals.  In particular, we reproduce the
$2\times 2$ Schwinger-Keldysh matrix propagator from a 
gravity calculation.  Our methods should work
also for computing higher point Lorentzian signature 
correlators.  We elucidate the boundary condition subtleties which
hampered previous efforts to build a Lorentzian-signature AdS/CFT
correspondence.  For two-point correlators, our construction is
automatically equivalent to the previously formulated prescription
for the retarded propagator.
\vfil
\begin{flushleft}
%November 2002
December 2002
\end{flushleft}
\vfil
\end{titlepage}
\newpage
\renewcommand{\baselinestretch}{1.1}  %looks better

%%%%%%%%%%%%%%%%%%%%%%%%%%%%%%%%%%%%%%%%%%%%%
%% include the next line for double spacing %%
%%%%%%%%%%%%%%%%%%%%%%%%%%%%%%%%%%%%%%%%%%%%%%
%\renewcommand{\baselinestretch}{2}
\renewcommand{\arraystretch}{1.5}

\section{Introduction}

The original AdS/CFT correspondence \cite{jthroat, EW, GKP}, 
motivated by considering 
a stack of D3-branes in ten dimensional
space, states that 
${\mathcal N}=4$ SU($N$) super Yang-Mills theory 
is dual to type IIB string theory in the background of 
five-dimensional anti-de Sitter space (AdS$_5$) cross
a five-sphere ($\bS^5$). 
Although the string theory
in this background is difficult to work with,
the low-energy supergravity (SUGRA) limit 
of string theory is accessible to quantitative calculations.
This limit corresponds to the regime of large 't Hooft coupling
in the gauge theory.
There exists a more general 
correspondence between conformal field theories
at finite temperature and asymptotically AdS spaces
containing black holes
\cite{Wittentherm}.  Noticing that the real-time formulation
of finite temperature field theory (the Schwinger-Keldysh,
or close-time-path, formalism)
involves a doubling of the degrees of freedom
\cite{Schwinger,Mahanthappa,Keldysh},  
and that the full Penrose diagram for asymptotically AdS
containing a black hole has two boundaries,
many have  
\cite{Maldacena, HorMar, vijay2} conjectured 
that the doubler fields can
be thought of as fields living on the
second boundary of the AdS dual. 

In this paper, we suggest a precise prescription for computing real-time
Green's functions from gravity.  We show how one can reproduce the full 
$2\times 2$ matrix
of two-point correlation functions
for a scalar field and its doubling partner
using the AdS dual (recall that in the real-time formalism, the 
mixed two-point 
correlators involving one real field
and one doubler field do not vanish).
Moreover, as our approach
is nothing more than a refinement of the
usual prescription of taking functional
derivatives of a boundary gravitational
action, the procedure should easily 
generalize to higher-point correlators.

There have been many previous attempts to
obtain Minkowski-signature correlation
functions from AdS/CFT.
The early prescriptions
for matching correlation functions in the field
theory to the classical behavior of bulk fields
in the supergravity involved a Wick
rotation to a Euclidean signature metric.
This rotation works only at zero temperature, and
perhaps obscures the point that
the correspondence should work equally
well for the original Minkowski signature.
The difficulty with working in the
original Minkowski signature is that
one generally has greater freedom to set
boundary conditions, and it is not clear a priori which boundary
condition corresponds to which propagator in a rather large set
of Green's functions:
advanced, retarded, Feynman, etc.  

Although the Euclidean correlation functions
can in principle be analytically continued
to yield the Feynman Green's function, and through it all
physical Minkowski signature
correlators, it is often desirable to
be able to compute the Minkowski signature
correlators directly.  
Direct computation eliminates the need for analytic continuation from
a discrete set of Matsubara frequencies which can be technically 
difficult.
Moreover, in
finite-temperature AdS/CFT, one can solve the bulk
field equation on the gravity side only in the 
large or small frequency limits.  Analytically 
continuing a function that is only known in
certain limits is not always possible.

Minkowski-signature prescriptions have been 
investigated before \cite{vijay2, vijay}, and
recently a concrete proposal has been put forward~\cite{SS}.  
Ref.~\cite{SS} makes a particular choice of
boundary conditions and identifies the retarded Green's function
$G_R$ in one particular term of the resulting
boundary gravitational action.
In (1+1)-dimensional conformal field theory where the Euclidean
Green's function can be computed exactly
and the analytic continuation to
a Minkowski signature metric can
be performed, the prescription gives the correct $G_R$~\cite{SS}.
Moreover, in the infrared limit the $G_R$ computed from the prescription
satisfies the constraints imposed by hydrodynamics
(see e.g. \cite{PSS, CH, PSS2}).
However, the prescription of Ref.~\cite{SS} does not follow from
taking functional derivatives, and so
the prescription cannot be directly generalized
to higher point correlators.

In contrast, the prescription presented in this paper allows for the
calculation of all correlators, including the higher-point ones, in
finite temperature field theories with asymptotically AdS (aAdS)
duals.  The zero-temperature correlators can be
obtained as a limit.
We will also see how the doubler fields of the 
Schwinger-Keldysh formalism emerge from gravity.

Our results come from understanding
black hole physics.  Our field theories
are dual to asymptotically AdS spaces containing
black holes, and we draw heavily on the ideas
of Hawking and Hartle \cite{HH}, Unruh \cite{Unruh},
and Israel \cite{Israel} who studied how black
holes produce thermal radiation.  
The choice of boundary condition for the
bulk fields in AdS are sensitive to the choice
of vacuum and coordinate system in ways that
are well studied in the context of Hawking radiation.
We will see that the thermal nature of black hole
physics gives rise to the thermal nature
of the field theory in a more or less
straightforward way.  

Saving the details for sections 3 and 4, 
note that the analog of Kruskal coordinates
exists for these aAdS spaces containing black
holes.  The correct prescription for
calculating Minkowski signature Green's functions
is the usual AdS/CFT
prescription worked out by \cite{EW, GKP} 
where one selects ``natural" boundary
conditions at the horizon with respect to
the analog of Kruskal time.  
In the context of Hawking radiation, Kruskal
time is often used to define  
a vacuum state.  With respect to the
gauge theory time, on the other hand, an observer
should see a thermal background.  If we
were to use gauge theory time to examine
the bulk behavior of a field, we would
need to take into account this thermal
background, a complication which
explains some of the early
confusion in the literature with
respect to these Minkowski signature
Green's functions.

In Kruskal
coordinates, the full Penrose diagram
(see figure 1) for aAdS space becomes
apparent.  Israel \cite{Israel} pointed
out that the fields in the mirror image universe
in the L quadrant of the Penrose diagram
should be the doubler fields of
the Schwinger-Keldysh formalism
in curved space.  The authors
of
\cite{Maldacena, HorMar, vijay2}
 made the further 
conjecture that the fields on the 
boundary of the L quadrant should look
like ghosts (or doubler fields) 
from the point of view
of the finite temperature CFT dual.
Being careful about boundary
conditions, we are able to
reproduce the $2 \times 2$ matrix
of propagators for the field and its doubler
using the aAdS description.

\begin{figure}[ht]
\begin{center}
\includegraphics[width=2.5in]{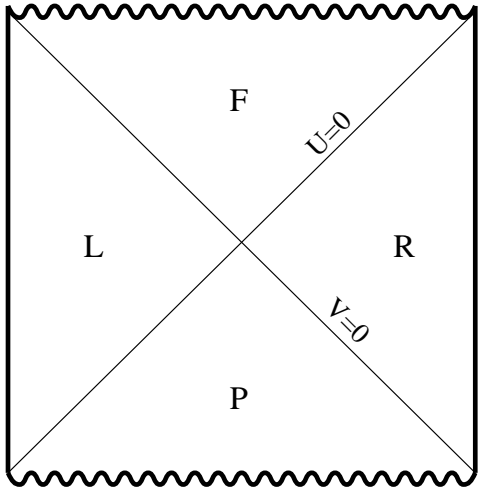}
\end{center}
\caption{The Penrose diagram for AdS containing a black hole.}
\label{fig1}
\end{figure}

We begin by reviewing in section two 
the Schwinger-Keldysh formalism
for real-time finite temperature field theory.  Section
three contains details of our refined prescription for 
calculating directly Minkowski signature correlators
in AdS/CFT.  We focus on the case of a scalar in
a non-extremal D3-brane background, but the prescription
should be much more broadly applicable.
In section 4, we explain how our choice of boundary
conditions are related to the boundary conditions
imposed by a Feynman propagator.

\section{Review of Schwinger-Keldysh Formalism for Finite-Temperature 
Field Theory}

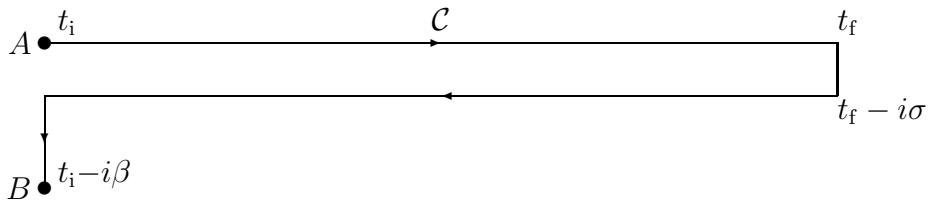
\begin{figure}[h]
\begin{center}
\begin{picture}(350,75)(-15,-5)
%\put(-15,-5){\dashbox{1}(350,75){}}
\put(0,55){\circle*{5}}
\put(0,55){\vector(1,0){150}}
\put(150,55){\line(1,0){150}}
\put(300,55){\line(0,-1){20}}
\put(300,35){\vector(-1,0){150}}
\put(150,35){\line(-1,0){150}}
\put(0,35){\vector(0,-1){18}}
\put(0,17){\line(0,-1){17}}
\put(0,0){\circle*{5}}
\put(5,58){\makebox(0,0)[bl]{$\ti$}}
\put(-5,55){\makebox(0,0)[r]{$A$}}
\put(-5,0){\makebox(0,0)[r]{$B$}}
\put(300,58){\makebox(0,0)[bl]{$\tf$}}
\put(150,60){\makebox(0,0)[b]{$\mathcal{C}$}}
\put(300,34){\makebox(0,0)[tl]{$\tf-i\sigma$}}
\put(5,0){\makebox(0,0)[bl]{$\ti{-}i\beta$}}

%\put(335,0){\line(0,1){125}}
\end{picture}
\end{center}
\caption{The Schwinger-Keldysh contour}
\label{fig:contour}
\end{figure}

In the Schwinger-Keldysh formalism, fields (which we denote
generically by $O$) live on a time contour $\mathcal{C}$ which goes
from some initial time $\ti$ to $\ti-i\beta$, but makes an excursion
along the real time axis in between.  A version of the contour is
drawn in Fig.~\ref{fig:contour}.  The contour starts at an initial
time $\ti$, goes to some (final) time $\tf$, then turns to the
Euclidean domain and runs to $\tf-i\sigma$ (where $\sigma$ is an
arbitrary length), after which it runs backward along the real time
axis to $\ti-i\sigma$ and then again turns to the Euclidean direction
and goes to $\ti-i\beta$.  The starting point $A$ (corresponding to
time $\ti$) and the ending point $B$ (corresponding to $\tf$) of the
contour are identified, and one requires that $O|_B=O|_A$ if $O$ is
bosonic and $O|_B=-O|_A$ if $O$ is fermionic.  For definiteness, we
will assume $O$ to be bosonic.

The parameter $\sigma$ can be chosen arbitrarily \cite{Matsu}. 
One possible choice
is $\sigma=0$, in which case the two Minkowski parts of the contour
lie on top of each other \cite{Schwinger, Keldysh}.  
To make contact with gravity another
choice, $\sigma=\beta/2$, is the most convenient.\footnote{
The $\sigma=\beta/2$ contour was studied 
in a field theoretic context in \cite{Matsu, NS}.}

The action of a field configuration is the sum of contributions from
the four parts of the contour,
\begin{equation} 
%\begin{split}
  S = \int\limits_{\cal C}\!dt\, L(t) = 
  \int\limits_\ti^\tf\!dt\, L(t)
  -i\!\int\limits_0^\sigma\!d\tau\, L(\tf-i\tau)
  -\int\limits_\ti^\tf\!dt\, L(t-i\sigma) 
  -i\!\int\limits_\sigma^\beta\!d\tau\, L(\ti-i\tau)\,,
%\end{split}
\end{equation}
where
\begin{equation}
  L(t) = \int\!d\x\, \mathcal{L}[\phi(t,\x)]\,,
\end{equation}
and $\mathcal{L}$ is the Lagrangian density.

By introducing a source which is not vanishing on the two Minkowski parts
of the contour, one defines the generating functional
\begin{equation}
  Z[\phi_1,\phi_2] = \int\!{\cal D}\phi\,\exp\left(iS 
  +i\!\int\limits_\ti^\tf\!dt\!\int\!d\x\,\phi_1(x)O_1(x)
  -i\!\int\limits_\ti^\tf\!dt\!\int\!d\x\,\phi_2(x)O_2(x) 
  \right) \ .
\end{equation}
Here $\phi_{1,2}$ and $O_{1,2}$ are the source and the fields on the
two Minkowski parts of the contour, i.e.,
\begin{subequations}
\begin{eqnarray}
  \phi_1(t,\x) = \phi(t,\x)\,, & \qquad & O_1(t,\x) = O(t,\x)\,,\\
  \phi_2(t,\x) = \phi(t-i\sigma,\x)\,, & \qquad & 
    O_2(t,\x) = O(t-i\sigma,\x)\,.
\end{eqnarray}
\end{subequations}

%When the source is turned off, $J_1=J_2=0$, the path integral is
%simply equal to the statistical partition function,
%\begin{equation}
%  Z[0,0] = \Tr e^{-(\beta-\sigma) H} e^{iH(\tf-\ti)} 
%   e^{-\sigma H} e^{-iH(\tf-\ti)}
%  = \Tr e^{-\beta H}\,.
%\end{equation}
By taking second variations of $Z$ with respect to the source $\phi$
one finds the Schwinger-Keldysh propagator,
\begin{equation}
  iG_{ab}(x-y) = 
  \frac 1{i^2}\,\frac{\delta^2\ln Z[\phi_1,\phi_2]}
  {\delta\phi_a(x)\,\delta\phi_b(y)}
   = i \left(\begin{array}{cc} G_{11} & -G_{12}\\-G_{21} & G_{22}
  \end{array}\right) \ .
\end{equation}

In the operator formalism, the Schwinger-Keldysh propagator
corresponds to the contour-ordered correlation function.  In contour
ordering, time is ordered normally on the upper part of the contour
and reversely on the lower part, and moreover any point on the lower
part is considered to have larger ``contour time'' than any point on
the upper part.  This means
\begin{equation}\label{Gab-oper}
\begin{split}
  iG_{11}(t,\x) = \< T O_1(t,\x)O_1(0)\>\,,\qquad & 
  iG_{12}(t,\x) = \< O_2(0)O_1(t,\x)\>\,,\\
  iG_{21}(t,\x) = \< O_2(t,\x)O_1(0)\>\,,\qquad &
  iG_{22}(t,\x) = \< \bar T O_2(t,\x)O_2(0)\>\,.
\end{split}
\end{equation}
where $\bar T$ denotes reversed time ordering, and
\begin{subequations}
\begin{eqnarray}
  O_1(t,\x) &=& e^{iHt-i\P\cdot\x} O(0) e^{-iHt+i\P\cdot\x}\,,\\
  O_2(t,\x) &=& e^{iH(t-i\sigma)-i\P\cdot\x} O(0) 
     e^{-iH(t-i\sigma)+i\P\cdot\x}\,.
\end{eqnarray}
\end{subequations}

The Schwinger-Keldysh correlators are related to the retarded and
advanced Green's functions, which are defined as
\begin{subequations}\label{GRA}
\begin{eqnarray}
  iG_R(x-y) &=&  \theta(x^0-y^0)\<[O(x),\,O(y)]\>\,,\\
  iG_A(x-y) &=& \theta(y^0-x^0)\<[O(y),\,O(x)]\>\,.
\end{eqnarray}
\end{subequations}
If one goes to momentum space,
\begin{equation}
  G(k) = \int\!dx\,e^{-ik\cdot x}G(x)\, ,
\end{equation}
we find that
\begin{equation}
  G_A(k) = G_R^*(k)\,,
\end{equation}
and, by inserting the complete set of states into the definitions
(\ref{Gab-oper}) and (\ref{GRA}), one finds the following relations
between the Schwinger-Keldysh correlators and the retarded one,
\ifthenelse{\haveamsmath=1}{%
\begin{subequations}
\begin{eqnarray}\label{SKprop}
  G_{11}(k) &=& \Re G_R(k) + i\coth\frac\omega{2T}\,\Im G_R(k)\,,
  \qquad \omega \equiv k^0\,,\\
  G_{12}(k) &=& \frac{2i e^{-(\beta-\sigma)\omega}}
    {1-e^{-\beta\omega}}\,\Im G_R(k)\,,\\
  G_{21}(k) &=& \frac{2i e^{-\sigma\omega}}
    {1-e^{-\beta\omega}}\,\Im G_R(k)\,,\\
  G_{22}(k) &=& -\Re G_R(k) + i\coth\frac\omega{2T}\,\Im G_R(k)\,.
\label{SKprop2}
\end{eqnarray}
\end{subequations}}{%
\begin{eqnarray}
  G_{11}(k) &=& \Re G_R(k) + i\coth\frac\omega{2T}\,\Im G_R(k)\,,
  \qquad \omega \equiv k^0\,, \nonumber \\
  G_{12}(k) &=& \frac{2i}{e^{\beta\omega}-1}\,\Im G_R(k)\,,
   \nonumber \\
  G_{21}(k) &=& \frac{2i}{1-e^{-\beta\omega}}\,\Im G_R(k)\,,
   \nonumber \\
  G_{22}(k) &=& -\Re G_R(k) + i\coth\frac\omega{2T}\,\Im G_R(k)\,.
\label{SKprop}
\end{eqnarray}}
One sees that when $\sigma=\beta/2$ the matrix $G_{ab}$ is symmetric,
$G_{12}=G_{21}$, which makes this choice convenient.  We will see that
this particular value of $\sigma$ appears naturally in gravity.

\section{Lorentzian Field Theory Correlators from Gravity}
\label{sec:core}

Our formalism for computing real-time Green's functions
should work for a broad class of finite-temperature field theories.
The theories are required to
have an asymptotically anti-de Sitter (aAdS) space dual with a 
Schwarzschild-type black hole in the center.  The zero-temperature
correlators can be obtained in the limit where the black hole
vanishes.
Typical
examples of such asymptotically AdS spaces arise from studying
non-extremal D3-branes (AdS$_5$), M2-branes (AdS$_4$), or M5-branes
(AdS$_7$).  AdS$_3$ arises in studying collections of 
non-extremal D1- and D5-branes. 

To illustrate the formalism with a concrete example,
we will focus on 
the case of non-extremal D3-branes.  We will
be using a gravitational description for calculating
correlators of ${\mathcal N}=4$ SU($N$) super Yang-Mills
theory at finite temperature in the $N \to \infty$ 
and $g_{YM}^2 N \to \infty$ limit.

The metric on a stack of
non-extremal D3-branes is
\be
ds^2 = H(r)^{-1/2} \left[ -f dt^2 + d\vec x^2 \right] + 
H(r)^{1/2} \left( f^{-1} dr^2 + r^2 d\Omega_5^2 \right) 
\ee
where $H(r) = 1+ R^4/r^4$, $f(r) = 1-r_0^4/r^4$,  
$d\Omega_5^2$ is the metric on a unit $\bS^5$, and 
$R^4 \sim N$ is proportional to the number of D3-branes.
In these coordinates, there is a horizon at $r=r_0$.
%and
%the boundary of the space is found in the limit $r \to \infty$.
In the near horizon limit ($r \ll R$), the metric becomes
\be
ds^2 = \frac{(\pi T R)^2}{u} \left( -f(u) dt^2 + d\vec x^2 \right)
+ \frac{R^2}{4u^2 f(u)} du^2 + R^2 d\Omega_5^2 \ 
\ee
where $T=r_0/\pi R^2$ is the Hawking temperature and we have
introduced $u=r_0^2/r^2$.  Here, $u=0$ is the boundary of this
aAdS space while $u=1$ corresponds to the horizon.

As a warm-up and review, let us consider the behavior of 
a scalar field $\phi$ of mass $m$ in this aAdS space.
The field $\phi$ obeys the wave equation.
\begin{eqnarray*}
0 &=& \frac{1}{\sqrt{-g}} \partial_\mu \sqrt{-g} g^{\mu \nu} \partial_\nu \phi
- m^2 \phi \\ 
&=& 4u^3 \partial_u \left( \frac{f}{u} \partial_u \phi \right)
+ \frac{u}{(\pi T)^2} \left( - \frac{1}{f} \partial_t^2 +
\partial_{\vec x}^2 \right)\phi - m^2 R^2 \phi 
\end{eqnarray*}
where the $\mu, \nu, \ldots$ run over the indices of the aAdS$_5$, 
$t$, $x^1$, $x^2$, $x^3$, and $u$.  To analyze this differential
equation, one makes a Fourier decomposition
\be
\phi(x,u) = \int\! \frac{d^4k}{(2\pi)^4}\,e^{i k \cdot x}\,\phi(k, u)   \ .
\ee

The function $\phi(k,u)$ satisfies the equation
\be 
0 = 4u^3 \partial_u \left( \frac{f}{u} \partial_u \phi(k,u) \right)
+\frac{u}{(\pi T)^2 f} \left(\omega^2 - f |\vec k|^2 \right)\phi(k,u) - 
m^2 R^2 \phi(k,u) 
\label{scalarwe}
\ee
which we cannot solve analytically.
However, we can analyze the solution at large and small $u$.  At small
$u$, close to the boundary, the space looks like ordinary 
$AdS_5$ and we get two possible scalings,
\be
\phi(k,u) \approx \phi(k) u^{\alpha_-} + A(k) u^{\alpha_+}
\ee
where $\alpha_\pm$ is a solution to $4 \alpha(\alpha-2) = m^2 R^2$.
Close to the horizon, 
$\phi \sim (1-u)^\beta$ where $\beta = \pm i \omega / 4\pi T$
and one can choose either sign.  In Euclidean signature, 
$\beta$ is real: $\beta=\pm\omega/4\pi T$ 
and only one solution is
well behaved close to $u=1$.  
%To introduce a bit
%of notation, we will denote by $f_k^i(u)$ the solutions
%$\phi(k,u)$ where $\beta = - i \omega / 4\pi T$ where
%the superscript $i$ stands for incoming.  The solutions with
%the other choice of sign will be labeled $f_k^o(u)$
%where $o$ stands for outgoing.

The AdS/CFT prescription fixes one boundary condition
at the boundary $u=0$ by 
%requiring that $\phi(k,u)$ is equal to some 
fixing $\phi(k)$. 
%$\phi(k)$ at small $u$.  
In
the Euclidean case, regularity of the solution at
$u=1$ provides the other boundary condition.
However, in Minkowski signature there
is an ambiguity, since now both solutions are regular
at the horizon. 
%and 
%the question is  
%what choice of boundary conditions at the
%horizon is physical and/or relevant
%to AdS/CFT correspondence.

The authors of Ref.~\cite{SS} take the point of view that
the black hole should absorb everything that
comes to the event horizon and so take purely incoming
boundary conditions.  
% For $\omega>0$, 
This statement would seem to mean 
$\phi(x,u) \sim e^{-i\omega t} (1-u)^{-i \omega/4\pi T}$
near the horizon.  As we mentioned, the prescription of Ref.~\cite{SS}
does not allow one to compute higher-point Green's functions; thus
we need a more general prescription.  To describe this prescription one
needs to use a coordinate system which covers the whole Penrose diagram.
This system is analogous to the Kruskal coordinates for 
Schwarzschild black holes.  Indeed, close to the horizon, 
our metric reduces to 
a Schwarzschild black hole, 
\be
ds^2 \to 2 (\pi T R)^2  
\left( 
-\left(1-\frac{2M}{\rho}\right) dt^2 + 
\left(1-\frac{2M}{\rho}\right)^{-1} d\rho^2 \right) + \ldots \ 
\ee
where $T = 1 / 8 \pi M$ and $u = 2M/\rho$.
Near the horizon, the transformation
to the Kruskal coordinates $U$ and $V$ is the same as for Schwarzschild
black holes,
\begin{eqnarray*}
U &=& - 4 M e^{-(t-r_*)/4M} \ , \\
V &=& 4M e^{(t+r_*)/4M} 
\end{eqnarray*}
where $r_* = \rho + 2M \ln|(\rho/2M) - 1|$, keeping in mind
that $\rho \approx 2M$ in this near horizon limit.  
Kruskal time is defined as $t_K \equiv U+V$ while
the radial coordinate can be thought of as 
$x_K \equiv V-U$.  

In these Kruskal coordinates, the structure of 
the full Penrose diagram for this 
aAdS space becomes apparent (see figure 1).  
In the initial discussion of a scalar in aAdS, we 
were working in the 
R quadrant where $U<0$ and $V>0$ but there are three other quadrants, two
of which have singularities.  The L quadrant where $U>0$ and $V<0$ will
be very important in what is to follow.
The Kruskal radial coordinate $x_K$ has been chosen so that its value
increases as we move from the left to the right of the Penrose diagram.

In what follows we will need to distinguish between incoming and
outgoing modes, and positive- and negative-frequency modes.  We
illustrate the distinctions in the example of four plane-wave
solutions with frequency $\pm\omega$, $\omega>0$,
\begin{subequations}\label{4modes}
\begin{eqnarray}
  e^{-i\omega U} &=& e^{-i\omega(t_K-x_K)/2}\label{e-U} \ ,\\
  e^{-i\omega V} &=& e^{-i\omega(t_K+x_K)/2}\label{e-V} \ ,\\
  e^{i\omega U} &=& e^{i\omega(t_K-x_K)/2}\label{e+U} \ ,\\
  e^{i\omega V} &=& e^{i\omega(t_K+x_K)/2}\label{e+V} \ .
\end{eqnarray}
\end{subequations}
The modes (\ref{e-U}) and (\ref{e+U}) are outgoing (the wave front
moves to larger $x_K$ as $t_K$ increases), while (\ref{e-V}) and
(\ref{e+V}) are incoming waves.  In general, any solution to the plane
wave equations can be decomposed into the sum of a function of $U$ and
a function of $V$.  The part depending on $U$ is a superposition of
(\ref{e-U}) and (\ref{e+U}) and is outgoing on the R quadrant, while
the part depending on $V$ is a superposition of (\ref{e-V}) and
(\ref{e+V}) is incoming on the same quadrant.  In our terminology the
notion of incoming and outgoing is switched on the L quadrant.  We can
say that the notion of time $t$ far from the horizon reverses
direction in the L quadrant.

The subtleties of defining positive- and negative-frequency modes are
well known in the context of quantization of fields on the
gravitational background of a black hole~\cite{BD}.  By using $t_K$ to
define the vacuum state of the quantum field, one finds that an
observer far from the horizon experiences a thermal bath of radiation.
This fact can be translated into a horizon boundary condition for our
prescription.

Among the four plane waves considered above, (\ref{e-U}) and
(\ref{e-V}) have positive frequency, and (\ref{e+U}) and (\ref{e+V})
are of negative frequency.  If one extends these mode functions to the
complex $U$ and $V$ planes, one sees that the positive-frequency modes
(\ref{e-U}) and (\ref{e-V}) are analytic in the lower half of the $U$
or $V$ planes, while the negative-frequency modes (\ref{e+U}) and
(\ref{e+V}) are analytic in the upper half-planes.  Since taking
superposition does not alter these analytic properties, one finds that
a solution to the wave equation is composed of only positive-frequency
modes if it is analytic in the lower $U$ and $V$ half-planes, and vice
versa.

For our purposes it will be useful to work in the original $t$ and $r$
coordinates.  It is simply a matter of convenience, since any
function of $t$ and $r$ can be written in terms of $U$ and $V$.  In the
original coordinates one can solve the wave equation separately in the
R and L quadrants and obtain one set of mode functions in each
quadrant,
\begin{equation}\label{modeRL}
u_{k, R, \pm} = 
\left\{ \begin{array}{ll}
e^{ik\cdot x} f_{\pm k} (r) & \mbox{in R} \\
0 & \mbox{in L}
\end{array}
\right. \; \; \; \; \; \; \; \; \; \; 
u_{k, L, \pm} = 
\left\{ \begin{array}{ll}
0 & \mbox{in R} \\
e^{ik\cdot x} f_{\pm k} (r) & \mbox{in L} 
\end{array}
\right. \ .
%
%u_{k, L, +} = 
%\left\{ \begin{array}{ll}
%e^{ik\cdot x} f_k (r) & \mbox{in L} \\
%0 & \mbox{in R}
%\end{array}
%\right. \; \; \; \; &&
%
%u_{k, L, -} = 
%\left\{ \begin{array}{ll}
%e^{ik\cdot x} f_{-k} (r) & \mbox{in L} \\
%0 & \mbox{in R}
%\end{array}
%\right. 
\end{equation}
Some explanation of the notation is needed.  In Eq.~(\ref{modeRL}) 
$k$ is a four-vector $(k^0,k^1,\ldots)$, so $k\cdot x =
-k^0 t + k^1 x^1 + \cdots$. 
The function $f_k$ is a solution  
to the scalar wave Eq.~(\ref{scalarwe}) which behaves like
$e^{ik^0 r_*}$ near the horizon.
We denote $\omega = k^0$.
The function $f_k(r)$ has two important properties: $f_{-k}^* = f_k$,
and $f_k$
is independent of the sign of $\vec k = (k^1, k^2, \ldots)$.
By definition, the R modes vanish in the L
quadrant and the L modes vanish in the R quadrant.  As $t$ and $x$ denote
two separate coordinate systems in the
R and L quadrants, we add a subscript to 
distinguish $t_L$ from $t_R$.   

The four modes defined in Eq.~(\ref{modeRL}), in principle, can be
expanded in terms of modes defined in the Kruskal
coordinates~(\ref{4modes}).  Without performing the explicit Fourier
transform, one notices that near the horizon $\ukrp$ and $\uklp$ are
functions of $U$, so they are outgoing modes.  Analogously $\ukrm$ and
$\uklm$ are functions of $V$ and hence are incoming waves.  The
modes~(\ref{modeRL}) contain, however, both positive- and
negative-frequency parts.  To separate modes with different signs of
frequency one defines, following
Unruh~\cite{Unruh,BD}, the linear combinations that mix modes on the
two quadrants,
\begin{subequations}\label{u1234}
\begin{eqnarray}
u_{1,k} &=& \ukrp + e^{-\omega/2T} \uklp \ , \\
u_{2,k} &=& \ukrp + e^{\omega/2T} \uklp \ , \\
u_{3,k} &=& \ukrm + e^{\omega/2T} \uklm \ , \\
u_{4,k} &=& \ukrm + e^{-\omega/2T} \uklm \ .
\end{eqnarray}
\end{subequations}
To ensure that
\begin{equation}
u_1 \sim U^{i\omega/2\pi T} \; \; \;  
\mbox{and} \; \;
u_4 \sim V^{-i\omega/2\pi T} \ 
\end{equation}
are continuous across the horizon, a branch cut must be placed
in the upper halves of the complex $U$ and $V$ planes.  
Thus, these modes 
are analytic in the lower halves of the complex $U$ and $V$ planes,
and are positive-frequency.  Similarly,
\begin{equation}
u_2 \sim \bar U^{i\omega/2\pi T} \; \; \;  
\mbox{and} \; \;
u_3 \sim \bar V^{-i\omega/2\pi T} \ 
\end{equation}
are analytic in the upper halves of the complex $U$ and $V$ planes and
carry negative frequencies.  The characteristics of the modes
$u_{i,k}$ can be summarized as follows:
\begin{equation*}
\begin{split}
  u_{1,k}: &\quad \textrm{outgoing, positive-frequency} \ ,\\
  u_{2,k}: &\quad \textrm{outgoing, negative-frequency} \ ,\\
  u_{3,k}: &\quad \textrm{incoming, negative-frequency} \ ,\\
  u_{4,k}: &\quad \textrm{incoming, positive-frequency} \ .
\end{split}
\end{equation*}

According to the AdS/CFT philosophy, the generating functional
$Z[\phi_1,\phi_2]$ can be found by evaluating the action of a solution
to the field equations, with a boundary condition such that
$\phi_{1,2}$ are the values of the field at the boundaries.  The
Penrose diagram has two boundaries, so we assume that our
field $\phi$ is equal to $\phi_1$ on the boundary of the R quadrant
and $\phi_2$ on the boundary of the L quadrant.  Since a general
solution is a superposition of four modes~(\ref{u1234}), one needs to
impose boundary conditions at the horizon to eliminate two of the four
modes.

Since our goal is to reproduce the Schwinger-Keldysh propagator, which
is defined with contour time ordering, it is natural
to impose the condition that positive frequency modes should be purely
ingoing at the horizon in the R quadrant while negative frequency
modes should be purely outgoing at the horizon in the R quadrant.
Indeed, in field theory the Feynman propagator $G_F(x-y)$ contains
only positive-frequency modes in the limit $x^0\to\infty$ and
negative-frequency modes in the opposite limit
$x^0\to-\infty$.\footnote{For the free propagator this comes from the
fact that the poles of $G_F$ as a function of the complex frequency
$\omega$ are located below the real axis for $\omega>0$ and above for
$\omega<0$.}  In the next section we will show that at zero
temperature one can arrive at this ``natural'' boundary condition
independently by a Wick
rotation from Euclidean space.  These boundary conditions select out
$u_2$ and $u_4$ as the only components that we can use to describe the
bulk behavior of a real scalar field, so
\begin{equation} 
  \phi(x, r) = \sum_k \alpha_k u_{2,k} + \beta_k u_{4,k} \ .
  \label{bulkfield}
\end{equation}

We now have enough boundary conditions to specify uniquely the
behavior of the scalar field.  By requiring that~(\ref{bulkfield})
approaches $\phi_{1,2}$ on the two boundaries,
we can solve for $\alpha_k$ and $\beta_k$.  The
result reads
\begin{subequations}
\begin{eqnarray}
&&\begin{split}\phi(k, r)|_R &=  \left((n+1) f_k^*(r_R) -n f_k(r_R) \right)
\phi_1(k) \\
&\quad + \sqrt{n(n+1)} \left(f_k(r_R) - f_k^*(r_R) \right) 
\phi_2(k) \ , 
\end{split}\label{bulk1} \\ 
&&\begin{split} \phi(k, r)|_L &=
\sqrt{n(n+1)} \left(f_k^*(r_L) - f_k(r_L) \right)
\phi_1(k)  \\  
&\quad +  \left((n+1) f_k(r_L) - n f_k^*(r_L) \right)
\phi_2(k) 
 \ ,
\end{split} \label{bulk2}
\end{eqnarray}
\end{subequations}
where $n \equiv (\exp(\omega/T) - 1)^{-1}$.  
The $f_k(r)$ are normalized such that $f_k(r_B)=1$ at
the boundary.  
In the above expressions,
we took the Fourier transform of $\phi(x, r)$ with respect to $x_R$
for the portion in the R quadrant while we took the Fourier transform
with respect to $x_L$ for $\phi(x, r)$ in the L quadrant.  

From
these two equations (\ref{bulk1}) and (\ref{bulk2}),  
it is straightforward to read off the 
bulk-to-boundary propagators.  The added complication is 
that with two source terms and two different space-time
(or bulk) regions, there are now four different propagators.  
For instance, the first term on the right hand side of
Eq.~(\ref{bulk1}) is the bulk-to-boundary propagator for 
the R boundary and the R bulk.  The second term in this
equation is the bulk-to-boundary propagator for the L boundary and 
the R bulk.  Eq.~(\ref{bulk2}) then gives the bulk-to-boundary
propagators for the L bulk.

Now we are finally ready to apply the standard recipe from AdS/CFT
correspondence for computing Green's functions.  The classical
boundary action is
\begin{equation}
\frac{K}{2} \int_R \sqrt{-g} g^{rr} \phi(-k, r) \partial_r \phi(k, r) 
\frac{d^4k}{(2\pi)^4} - 
\frac{K}{2} \int_L \sqrt{-g} g^{rr} \phi(-k, r) \partial_r \phi(k, r) 
\frac{d^4k}{(2\pi)^4} \ ,
\end{equation}
where $K$ is some overall normalization.
The conjecture of Ref.~\cite{SS} is that the retarded and advanced
Green's functions are related to $f_k$ in the following way,
\begin{equation}\label{GRAf}
G_R(k) = -K \sqrt{-g} g^{rr} f_k(r) \partial_r f_k^*(r) |_{r_B} \; ;
\; \; \;
G_A(k) = -K \sqrt{-g} g^{rr} f_k^*(r) \partial_r f_k(r) |_{r_B} \ .
\end{equation}
Using the normalization of the $f_k$,
the radial derivative of $\phi(k, r)$ evaluated close to the R or L boundary is then
\begin{subequations}
\begin{eqnarray}
-K \sqrt{-g} g^{rr} \partial_r
\phi |_R &=& [(1+n) G_R - n G_A] \phi_1 + \sqrt{n(1+n)} 
   (G_A - G_R) \phi_2 \ , \\
-K \sqrt{-g} g^{rr} \partial_r
\phi |_L &=& [(1+n) G_A - n G_R] \phi_2 + \sqrt{n(1+n)} (G_R - G_A) \phi_1 \ .
\end{eqnarray}
\end{subequations}

The boundary action becomes
\begin{equation}
\begin{split}
S &= -\frac{1}{2} \int \frac{d^4k}{(2\pi)^4} \Bigl[ \phi_1(-k) 
\left( (1+n) G_R(k) - n G_A(k) \right) \phi_1(k) \\
&\quad
-\phi_2(-k) \left( (1+n) G_A(k) - n G_R(k) \right) \phi_2(k) \\
&\quad
 +\phi_1(-k) \sqrt{n(1+n)} (G_A(k) - G_R(k) ) \phi_2(k)  \\
&\quad
 + \phi_2(-k) \sqrt{n(1+n)} (G_A(k) - G_R(k) ) \phi_1(k) \Bigr] \ .
\end{split}
\end{equation}

Taking functional derivatives of $S$ with respect to $\phi_1(k)$ and
$\phi_2(k)$ yields precisely the Schwinger-Keldysh propagators
(\ref{SKprop})--(\ref{SKprop2})
with $\sigma=\beta/2$.  One thus concludes 
that if the
conjecture~(\ref{GRAf}) is valid, then the Schwinger-Keldysh
correlators can be found by taking functional derivatives of the
classical action.  Vice versa, if one take the classical action as the
starting point, then by taking functional derivatives one can find
all Schwinger-Keldysh correlators which, in conjunction with
Eqs.~(\ref{SKprop})--(\ref{SKprop2}) 
will give us the same retarded and advanced
Green's functions as computed from the old prescription of
Ref.~\cite{SS}.\footnote{
For more complicated, composite operators, such as the 
stress-energy tensor, we expect there may be additional
complications arising from contact terms.  
In particular, we believe that our boundary conditions
will continue to produce the Schwinger-Keldysh propagators.
However, the relation between $G_R$ and $f_k$ in 
Eq.~(\ref{GRAf}) 
and the relation between $G_R$ and $G_{ij}$
in Eqs.~(\ref{SKprop})--(\ref{SKprop2}) 
may change by contact terms.}

In order to obtain the Schwinger-Keldysh propagator with 
$\sigma\neq \beta/2$, one can 
substitute the source $\phi_2(k)$ in the boundary
action with $e^{(\sigma-\beta/2)\omega} \phi_2(k)$.
The interpretation of this rescaling of $\phi_2(k)$
is not completely clear from the gravity point of view.
%This substitution can be understood in terms of
%a shift in the definition of $t_L$.  
%treat the time variables on the 
%left and the
%right quadrants as different by $i\beta/2$: $t_R-t_L=-i/2T$.  Then
%Eq.~(\ref{bulk2}) should be multiplied by $e^{-\omega/2T}$.  One can
%see that the derivatives of the action with respect to the boundary
%fields reproduce Eqs.~(\ref{SKprop})--(\ref{SKprop2}) 
with $\sigma=0$.

\section{Natural boundary conditions at zero temperature}

In zero-temperature field theory one can perform a Wick rotation from
Euclidean space to Minkowski space.  We now show that the ``natural'' 
boundary condition at the horizon proposed in the previous 
section is consistent with this Wick rotation.
Consider the zero temperature limit
of aAdS space, 
where we recover a scalar traveling in the Poincare patch
of pure $AdS_5$.  In Euclidean signature, the 
behavior of the scalar field is described by
\be
f^E_k(z) = \frac{z^2 K_\nu(kz)}{\epsilon^2 K_\nu(k\epsilon)}
\ee
where $\nu = \sqrt{4+m^2R^2}$, $k = \sqrt{\omega^2 + (k^1)^2 + \cdots}$,
and $z=0$ corresponds to the boundary of $AdS_5$.\footnote{
Our metric is the usual 
\[
ds^2 = (\pm dt^2 + d\vec x^2 + dz^2)/z^2 \ .
\]}
The analytic continuation of the Bessel type function $K_\nu$ to
Lorentzian signature is the first Hankel function
$H_\nu^{(1)}$ and one finds that in Lorentzian signature
\be
f_k(z) = \frac{z^2 H_\nu^{(1)}(qz)}{\epsilon^2 H_\nu^{(1)}(q\epsilon)}
\ee
where $q= \sqrt{\omega^2-(k^1)^2-\cdots}$.  When multiplied by
$e^{-i\omega t}$, for large $z$, $f_k(z)$ corresponds to
a wave traveling away from the boundary for $\omega>0$ while
for $\omega<0$, the wave travels toward the boundary.  These
boundary conditions are precisely the zero temperature
limit of our ``natural'' boundary conditions.

The boundary conditions
on the Feynman propagator are the same conditions that result from
an analytic continuation of the Euclidean Green's function.
This zero temperature calculation tells us that at least at
zero temperature, the right boundary conditions are purely
outgoing from the boundary for positive frequency and purely
ingoing at the boundary for negative frequency modes.

Generalizing now to finite temperature, we are asserting
that the boundary conditions should remain the same but we
should use the Kruskal coordinates to define them and not
the original gauge theory time $t$.  

\section*{Acknowledgments}

The authors thank 
Oliver DeWolfe, Jim Hartle, 
Gary Horowitz,
Thomas Hertog, Joe Polchinski, Andrei Starinets,
Anastasia Volovich, and Johannes Walcher for discussions.
C.~H. was supported in 
part by the National Science Foundation under
Grant No. PHY99-07949.  D.~T.~S. was supported, in part, by DOE grant
No.\ DOE-ER-41132 and the Alfred P.\ Sloan Foundation.

\end{document}